\pdfoutput=1
\documentclass{PoS}

\usepackage{slashed}
\usepackage{amsfonts,amsmath,bm}

\title{Form factors for $\Lambda_b \to \Lambda$ transitions from lattice QCD}

\ShortTitle{Form factors for $\Lambda_b \to \Lambda$ transitions from lattice QCD}

\author{William Detmold$^{a}$, C.-J. David Lin$^{b,c}$, \speaker{Stefan Meinel}$\:^a$, Matthew Wingate$^d$\\ \\
\llap{$^a$}Center for Theoretical Physics, Massachusetts Institute of Technology, Cambridge, MA 02139, USA\\
\llap{$^b$}Institute of Physics, National Chiao-Tung University, Hsinchu 300, Taiwan\\
\llap{$^c$}Physics Division, National Centre for Theoretical Sciences, Hsinchu 300, Taiwan\\
\llap{$^d$}DAMTP, University of Cambridge, Wilberforce Road, Cambridge CB3 0WA, UK\\
\\ \\
E-mail: \email{smeinel@mit.edu}}

\abstract{The rare baryonic decays $\Lambda_b \to \Lambda \mu^+ \mu^-$ and $\Lambda_b \to \Lambda \gamma$
can complement rare $B$ meson decays in constraining models of new physics. In this work, we calculate
the relevant $\Lambda_b \to \Lambda$ transition form factors at leading order in heavy-quark effective theory
using lattice QCD. Our analysis is based on RBC/UKQCD gauge field ensembles with 2+1 flavors
of domain-wall fermions, and with lattice spacings of $a\approx 0.11$ fm and $a\approx 0.08$ fm.
We compute appropriate ratios of three-point and two-point correlation functions for a wide range
of source-sink separations,  and extrapolate to infinite separation in order to eliminate excited-state
contamination. We then extrapolate the form factors to the continuum limit and to the physical values of
the light-quark masses.}

\FullConference{The 30th International Symposium on Lattice Field Theory - Lattice 2012,\\
June 24-29, 2012\\
Cairns, Australia}

\begin{document}

\section{Introduction}

Flavor-changing neutral-current $b \to s$ decays play an important role in constraining models of new physics.
In addition to the widely-studied mesonic decays $B \to K^* \gamma$ and $B \to K^{(*)} \ell^+\ell^-$,
baryonic $b \to s$ decays decays such as $\Lambda_b \to \Lambda \gamma$ and $\Lambda_b \to \Lambda \ell^+ \ell^-$
are also worth investigating since the spin of the baryons allows the construction of observables that are sensitive
to the helicity structure of the effective weak Hamiltonian \cite{Mannel:1997xy}. The CDF collaboration
has recently observed the decay $\Lambda_b \to \Lambda \mu^+ \mu^-$ for the first time \cite{Aaltonen:2011qs},
and new results are expected from LHCb.

To calculate the decay amplitudes for $\Lambda_b \to \Lambda \gamma$ and $\Lambda_b \to \Lambda \ell^+ \ell^-$,
the hadronic matrix elements $\langle \Lambda(\mathbf{p'}, s') | \:\bar{s} \Gamma b\: | \Lambda_b(\mathbf{p}, s) \rangle$
need to be determined for
$\Gamma \in \{ \gamma_\mu,\: \gamma_\mu\gamma_5,\: q^\nu\sigma_{\mu\nu},\: q^\nu\sigma_{\mu\nu}\gamma_5 \}$ (where $q=p-p'$),
resulting in ten independent form factors. The situation simplifies when using heavy-quark effective theory (HQET)
for the $b$ quark. At leading order in heavy-quark effective theory, only two independent form factors remain,
and one has \cite{Mannel:1990vg}
\begin{equation}
\langle \Lambda(\mathbf{p'}, s') | \:\bar{s} \Gamma Q \: | \Lambda_Q(v, s) \rangle
= \overline{u}(\mathbf{p'},s')\left[ F_1(p'\cdot v) + \slashed{v}\:F_2(p'\cdot v) \right] \Gamma\: \mathcal{U}(v, s). \label{eq:FFdef}
\end{equation}
Here, $v$ is the four-velocity of the $\Lambda_Q$, and the form factors $F_1$ and $F_2$ are functions
of $p'\cdot v$, which is equal to the energy of the $\Lambda$ baryon in the $\Lambda_Q$ rest frame.
In our analysis, it proves more convenient to work with the linear combinations
\begin{equation}
 F_+ = F_1 + F_2, \hspace{8ex} F_- = F_1 - F_2, \label{eq:pm}
\end{equation}
instead of $F_1$ and $F_2$. In the following, we report on our calculation of these two form factors
using lattice QCD. For the static heavy quark $Q$, we set $v=(1,0,0,0)$, and we use a lattice HQET action
with one iteration of HYP smearing for the gauge link in the time derivative \cite{Eichten:1989kb}.
For the up, down, and strange quarks, we use a domain-wall action \cite{Kaplan:1992bt}. Our calculations
are based on the 2+1 flavor RBC/UKQCD gauge field ensembles described in Ref.~\cite{Aoki:2010dy}.

\section{Extracting the form factors from correlation functions}

In our two-point and three-point correlation functions, we use the following interpolating fields
for the $\Lambda_Q$ and $\Lambda$ baryons,
\begin{equation}
 \Lambda_{Q\alpha} = \epsilon^{abc}\:(C\gamma_5)_{\beta\gamma}
 \:\tilde{d}^a_\beta\:\tilde{u}^b_\gamma\: Q^c_\alpha, \hspace{8ex}
 \Lambda_{\alpha} = \epsilon^{abc}\:(C\gamma_5)_{\beta\gamma}
 \:\tilde{u}^a_\beta\:\tilde{d}^b_\gamma\: \tilde{s}^c_\alpha, \label{eq:Lambdainterpol}
\end{equation}
where the tilde on the up, down, and strange-quark fields indicates gauge-covariant Gaussian smearing.
In the three-point functions, we use an $\mathcal{O}(a)$-improved discretization of the continuum HQET
current, which is given by \cite{Ishikawa:2011dd}
\begin{equation}
 J_{\Gamma}^{(\rm HQET)}(m_b) = U(m_b, a^{-1})\: \mathcal{Z} \left[ \overline{Q}\: \Gamma \: s
 + c^{(m_s a)}_\Gamma\:\frac{m_s\:a}{1-(w_0^{\rm MF})^2}\: \overline{Q}\: \Gamma \: s 
 + c^{(p_s a)}_\Gamma\:a\: \overline{Q}\: \Gamma\: \bm{\gamma}\cdot \bm{\nabla} \: s   \right]. \label{eq:LHQETcurrent}
\end{equation}
The matching factor $\mathcal{Z}$ and the improvement coefficients $c^{(m_s a)}_\Gamma$ and $c^{(p_s a)}_\Gamma$
have been computed in one-loop lattice perturbation theory in Ref.~\cite{Ishikawa:2011dd}. The factor $U(m_b, a^{-1})$
provides two-loop running in continuum HQET from $\mu=a^{-1}$ to $\mu=m_b$.

\begin{figure}
\hfill \includegraphics[height=4.5cm]{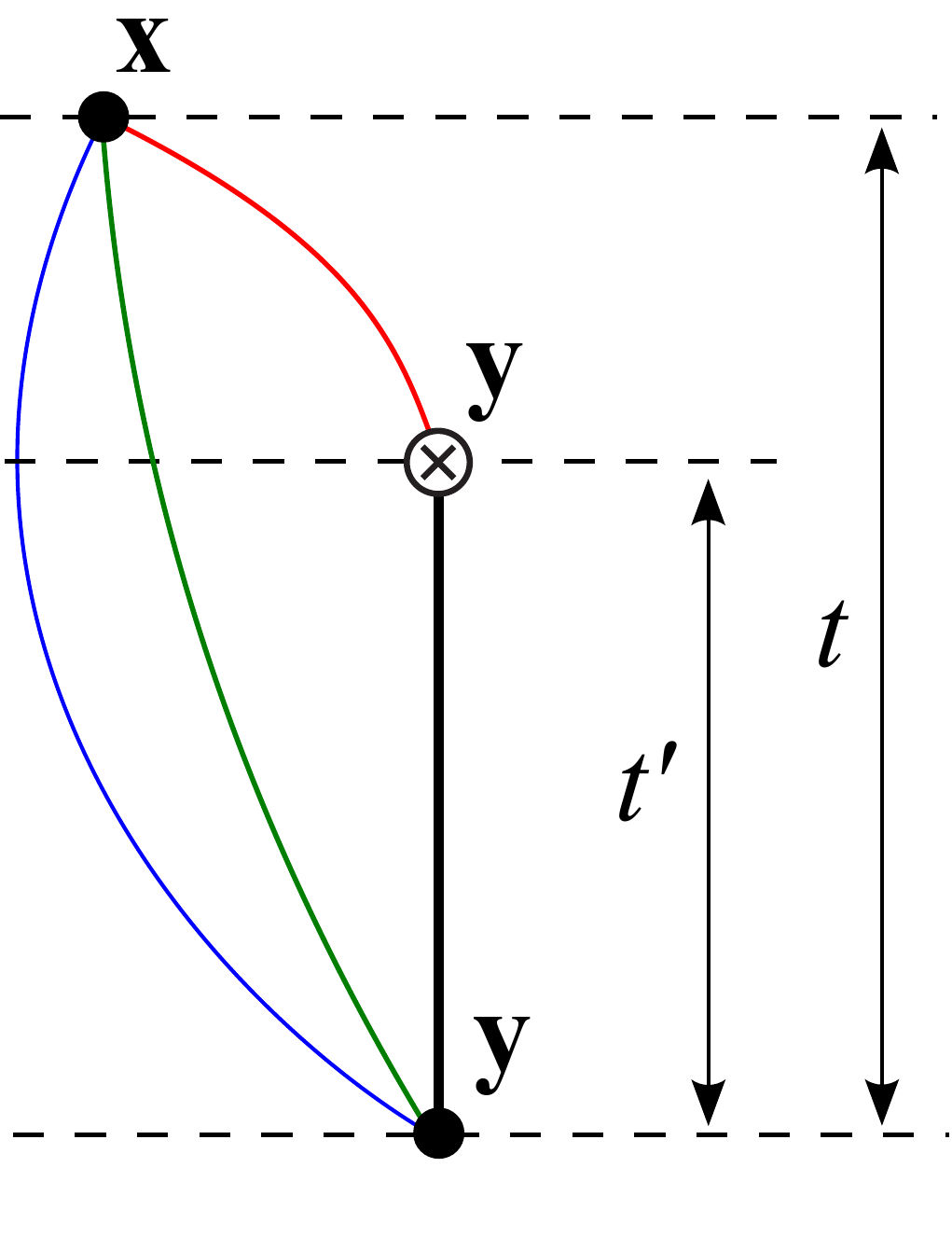}\hfill\includegraphics[height=4.5cm]{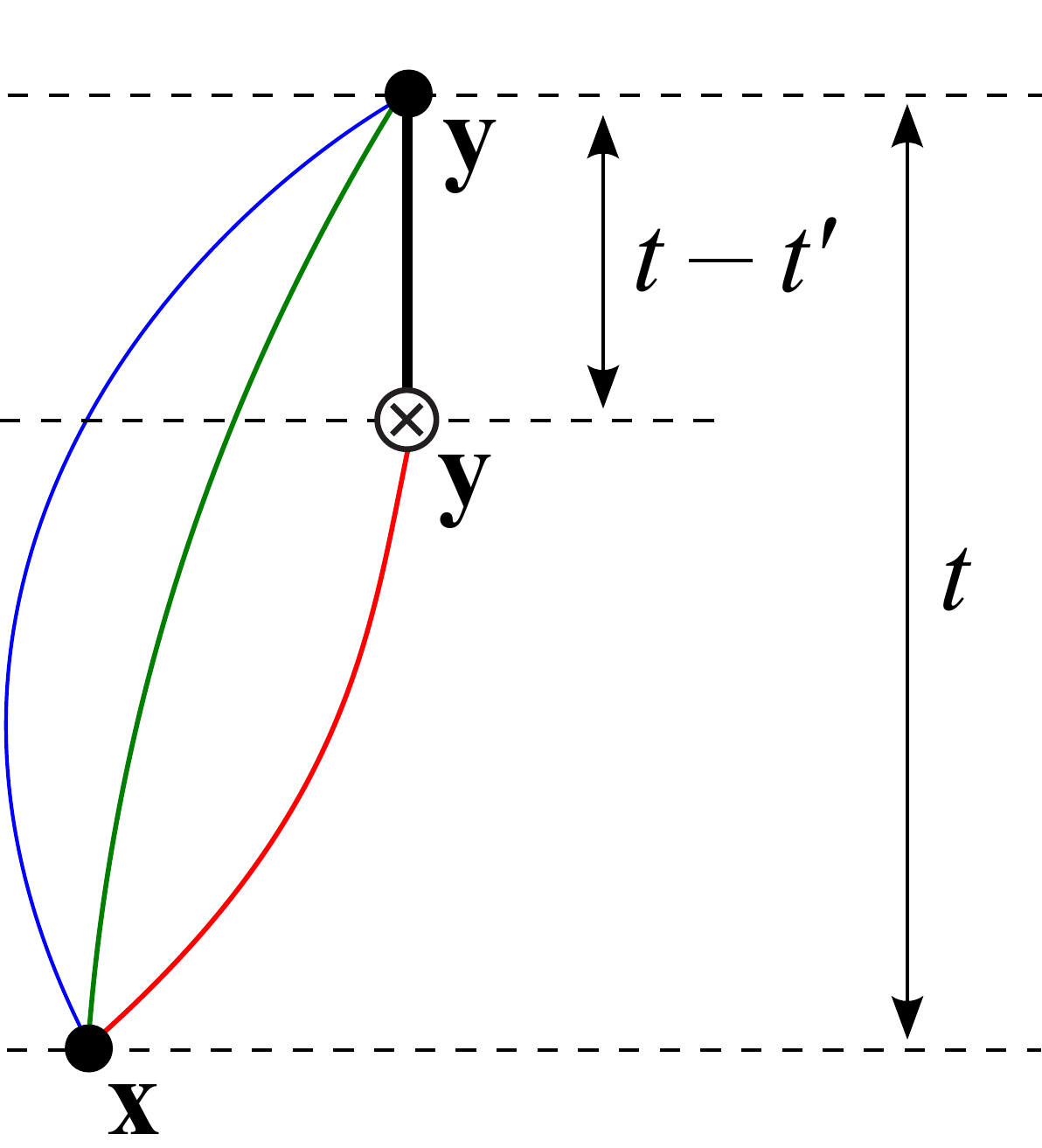} \hfill \null
\caption{\label{fig:threept}Propagator contractions for $ C^{(3)}_{\delta\alpha}(\Gamma,\:\mathbf{p'}, t, t')$ (left)
and $C^{(3,\mathrm{bw})}_{\alpha\delta}(\Gamma,\:\mathbf{p'}, t, t-t')$ (right).
The vertical thick lines indicate the static heavy-quark propagators.}
\end{figure}

We compute ``forward'' and ``backward'' three-point functions originating from a common source point $(x_0,\mathbf{x})$,
\begin{eqnarray}
 C^{(3)}_{\delta\alpha}(\Gamma,\:\mathbf{p'}, t, t') &=& \sum_{\mathbf{y}} e^{-i\mathbf{p'}\cdot(\mathbf{x}-\mathbf{y})}
 \left\langle \Lambda_{\delta}(x_0,\mathbf{x})\:\: J_\Gamma^{(\rm HQET)\dag}(x_0-t+t',\mathbf{y})
 \:\:\: \overline{\Lambda}_{Q\alpha} (x_0-t,\mathbf{y}) \right\rangle, \hspace{4ex} \label{eq:threept} \\
C^{(3,\mathrm{bw})}_{\alpha\delta}(\Gamma,\:\mathbf{p'}, t, t-t') &=& \sum_{\mathbf{y}}
e^{-i\mathbf{p'}\cdot(\mathbf{y}-\mathbf{x})} \left\langle \Lambda_{Q\alpha}(x_0+t,\mathbf{y})\:\: J_\Gamma^{(\rm HQET)}(x_0+t',\mathbf{y})
\:\:\: \overline{\Lambda}_{\delta} (x_0,\mathbf{x}) \right\rangle. \label{eq:threeptbw}
\end{eqnarray}
As is apparent from Fig.~\ref{fig:threept}, the three-point functions do not require sequential domain-wall propagators
and can be computed efficiently for arbitrary values of $t$, $t'$, $\Gamma$, and $\mathbf{p'}$. We then construct the ratio
\begin{equation}
 \mathcal{R}(\Gamma, \mathbf{p'}, t, t') = \frac{4 \:\mathrm{Tr}\left[  C^{(3)}(\Gamma,\:\mathbf{p'}, t, t')
 \:\: C^{(3,\mathrm{bw})}(\Gamma,\:\mathbf{p'}, t, t-t') \right] }{ \mathrm{Tr}[ C^{(2,\Lambda)}(\mathbf{p'}, t)]
 \:\mathrm{Tr}[ C^{(2,\Lambda_Q)}(t) ] }, \label{eq:doubleratio}
\end{equation}
where $C^{(2,\Lambda)}(\mathbf{p'}, t)$ and $C^{(2,\Lambda_Q)}(t)$ are the $\Lambda$ and $\Lambda_Q$ two-point functions.
By inserting complete sets of states and using Eq.~(\ref{eq:FFdef}), one finds that, for $\Gamma$ equal to any product
of $\gamma^\mu$'s,
\begin{equation}
 \mathcal{R}(\Gamma, \mathbf{p'}, t, t') = \left\{ \begin{array}{ll}
 \displaystyle\frac{ E_\Lambda+m_\Lambda}{E_\Lambda} [F_+]^2 + \hdots, &\hspace{2ex} {\rm if}\:\:\:[\Gamma,\gamma^0]=0, \\
 \displaystyle\frac{ E_\Lambda-m_\Lambda}{E_\Lambda} [F_-]^2 + \hdots, &\hspace{2ex} {\rm if}\:\:\:\{\Gamma,\gamma^0\}=0.
 \end{array} \right. \label{eq:RGamma}
\end{equation}
Here, $F_\pm$ are the form factors defined in Eq.~(\ref{eq:pm}), and the ellipsis indicates excited-state contributions
that decay exponentially with the time separations.  To increase statistics, we average the ratio over multiple gamma
matrices and define
\begin{eqnarray}
 \mathcal{R}_+(\mathbf{p'}, t, t') &=&
 \frac14 \left[ \mathcal{R}(1, \mathbf{p'}, t, t') + \mathcal{R}(\gamma^2\gamma^3, \mathbf{p'}, t, t')
 + \mathcal{R}(\gamma^3\gamma^1, \mathbf{p'}, t, t') + \mathcal{R}(\gamma^1\gamma^2, \mathbf{p'}, t, t') \right], \hspace{4ex}  \label{eq:curlyRplus} \\
 \mathcal{R}_-(\mathbf{p'}, t, t') &=&
 \frac14 \left[ \mathcal{R}(\gamma^1, \mathbf{p'}, t, t') + \mathcal{R}(\gamma^2, \mathbf{p'}, t, t')
 + \mathcal{R}(\gamma^3, \mathbf{p'}, t, t') + \mathcal{R}(\gamma_5, \mathbf{p'}, t, t') \right].  \label{eq:curlyRminus}
\end{eqnarray}
(Replacing $\Gamma$ by $\Gamma\gamma_0$ would not give new information because $\gamma_0 Q = Q$.)
For a given value of $|\mathbf{p'}|^2$, we then average $\mathcal{R}_\pm(\mathbf{p'}, t, t')$ over the direction
of $\mathbf{p'}$, and we denote these direction-averaged quantities as $\mathcal{R}_\pm (|\mathbf{p'}|^2, t, t')$.
Because of the symmetric form of our ratio (\ref{eq:doubleratio}), at a given source-sink separation $t$, the
excited-state contamination will be smallest at $t'=t/2$. We therefore define the new quantities
\begin{eqnarray}
 R_+(|\mathbf{p'}|^2, t) &=& \sqrt{\frac{E_\Lambda}{E_\Lambda+m_\Lambda} \mathcal{R}_+(|\mathbf{p'}|^2,\: t,\: t/2)}, \label{eq:Rplus}\\
 R_-(|\mathbf{p'}|^2, t) &=& \sqrt{\frac{E_\Lambda}{E_\Lambda-m_\Lambda} \mathcal{R}_-(|\mathbf{p'}|^2,\: t,\: t/2)}, \label{eq:Rminus}
\end{eqnarray}
which, according to Eq.~(\ref{eq:RGamma}), are equal to the form factors $F_\pm(v\cdot p')$ up to excited-state
effects that decay exponentially with the source-sink separation $t$. We obtain the energy $E_\Lambda(|\mathbf{p'}|^2)$
and mass $m_\Lambda$ appearing in Eqs.~(\ref{eq:Rplus}) and (\ref{eq:Rminus}) from fits
to the $\Lambda$ two-point functions for the same data set.

\section{Data analysis}

\begin{figure}
\includegraphics[height=4.5cm]{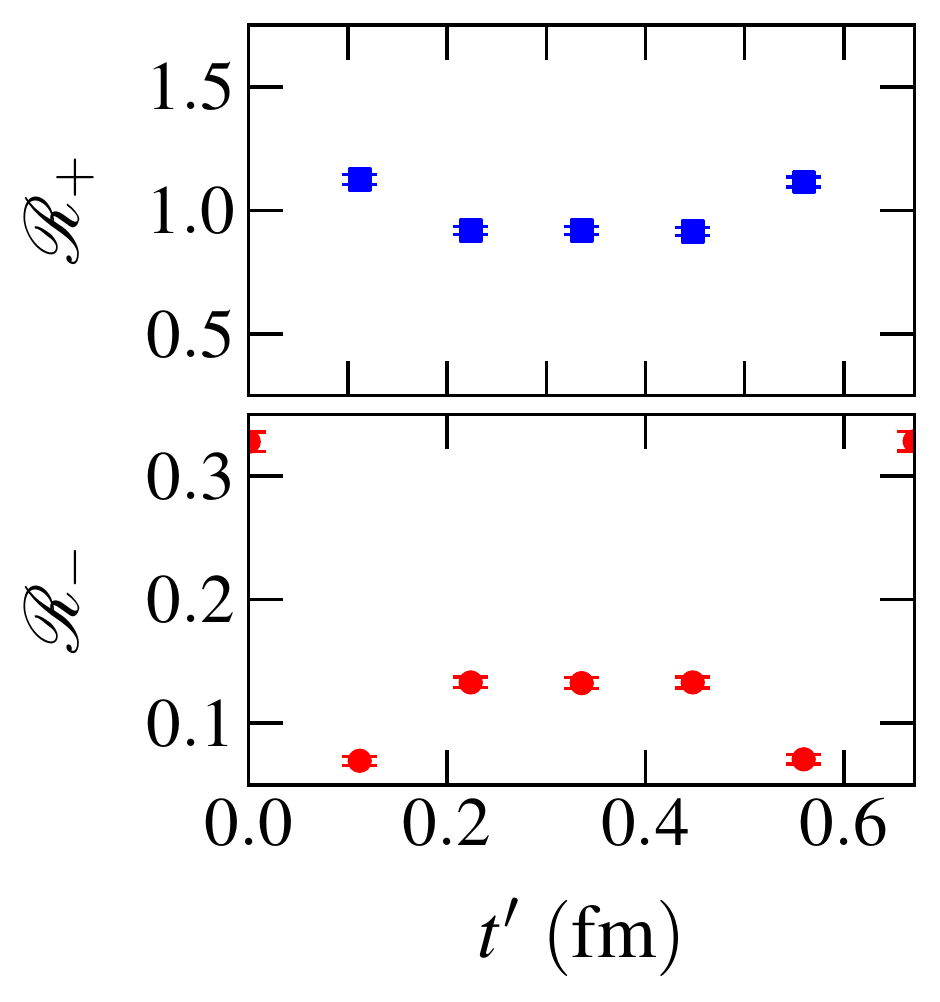}
\includegraphics[height=4.5cm]{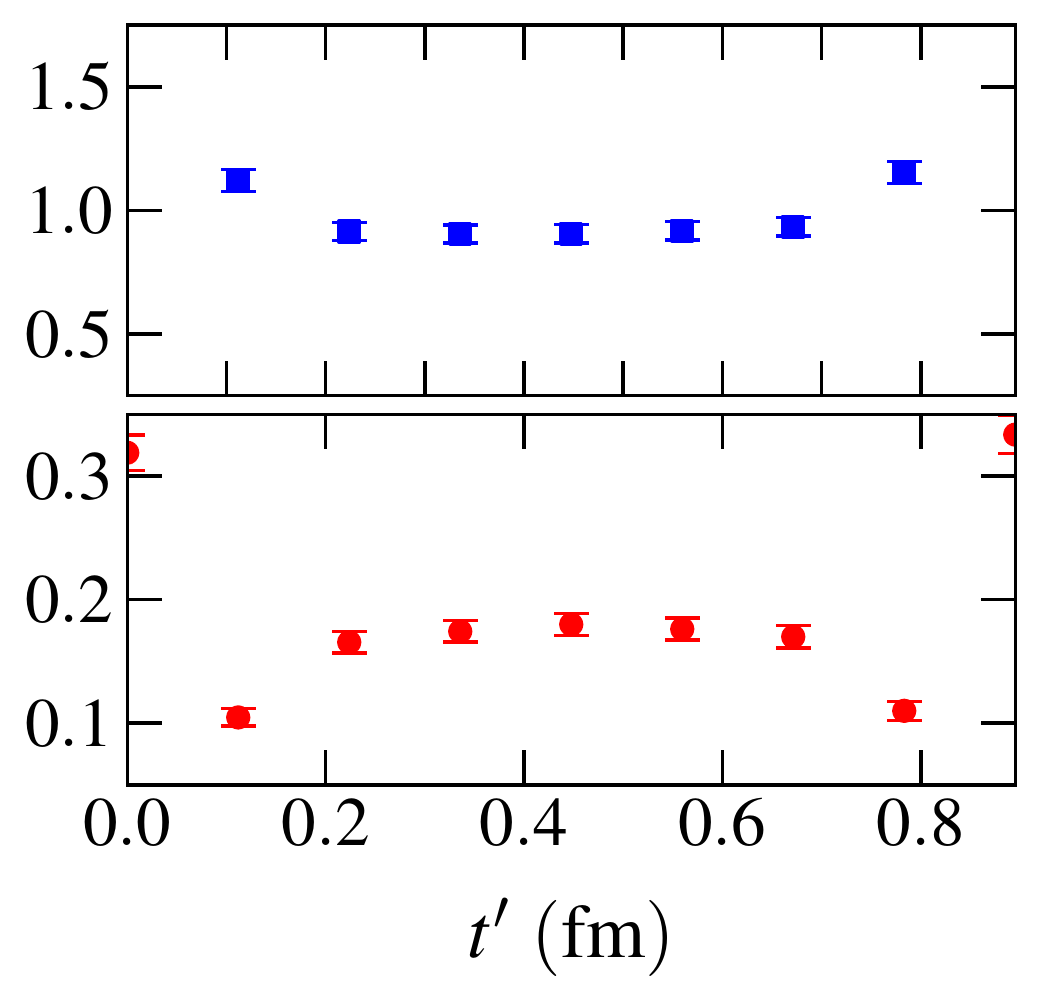}
\includegraphics[height=4.5cm]{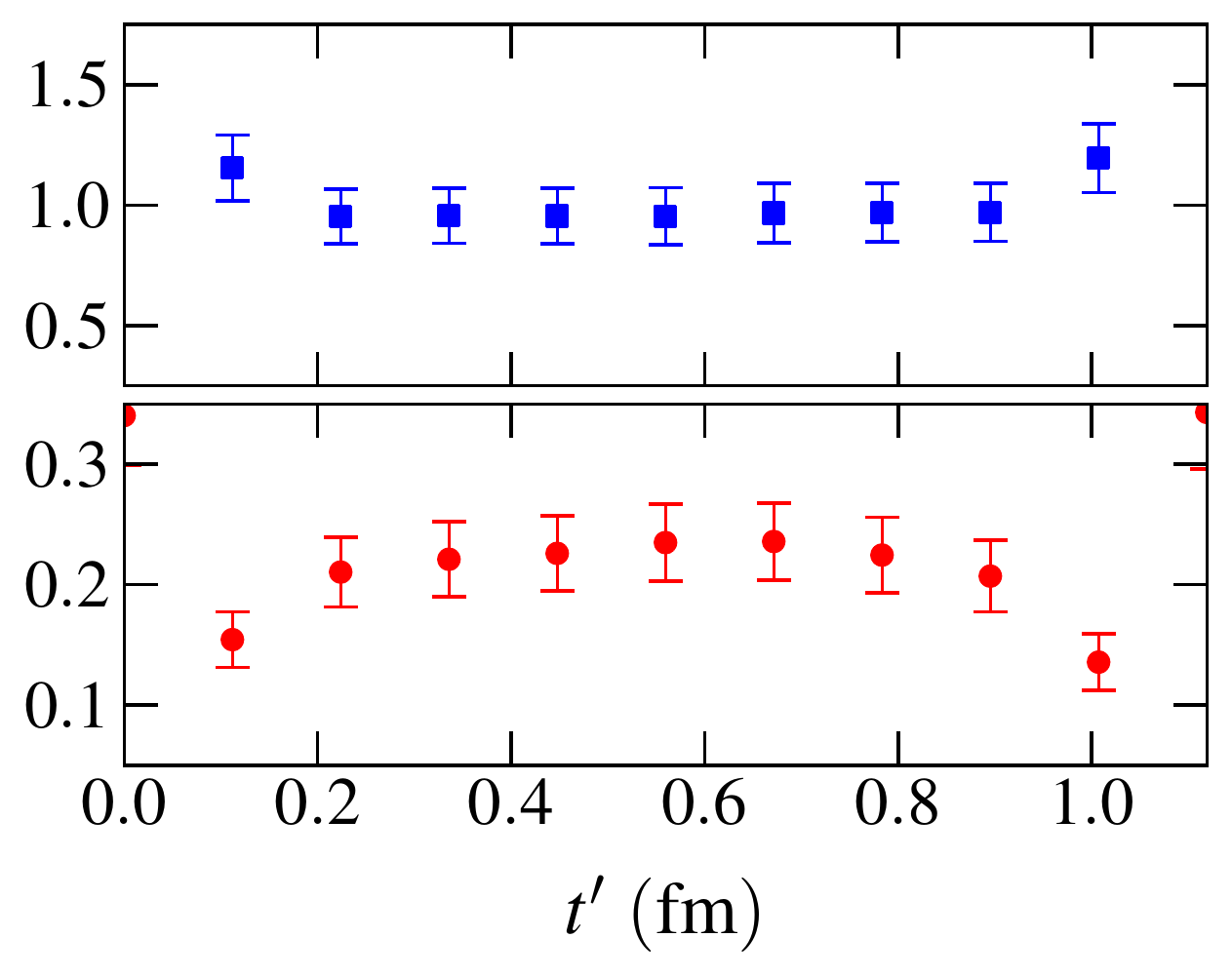}
\caption{\label{fig:ratios_L24_005_psqr4}Example results for the ratios $\mathcal{R}_\pm (|\mathbf{p'}|^2, t, t')$
from the $\mathtt{C54}$ data set. The data shown here are for $|\mathbf{p'}|^2=4\cdot(2\pi/L)^2$,
at source-sink separations (from left to right) $t/a=6,8,10$.}
\end{figure}

\begin{figure}
\includegraphics[width=0.48\linewidth]{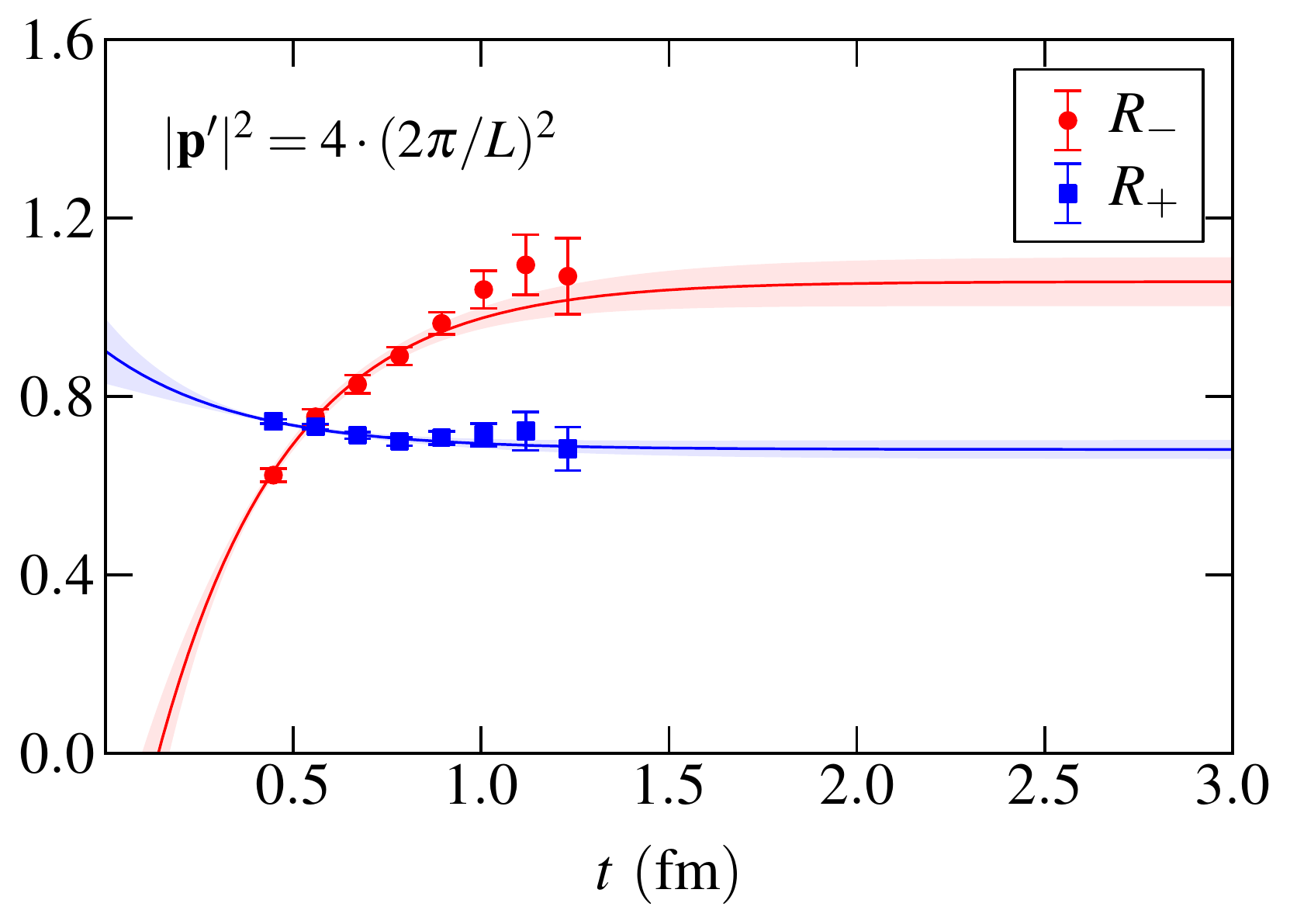}
\hfill\includegraphics[width=0.48\linewidth]{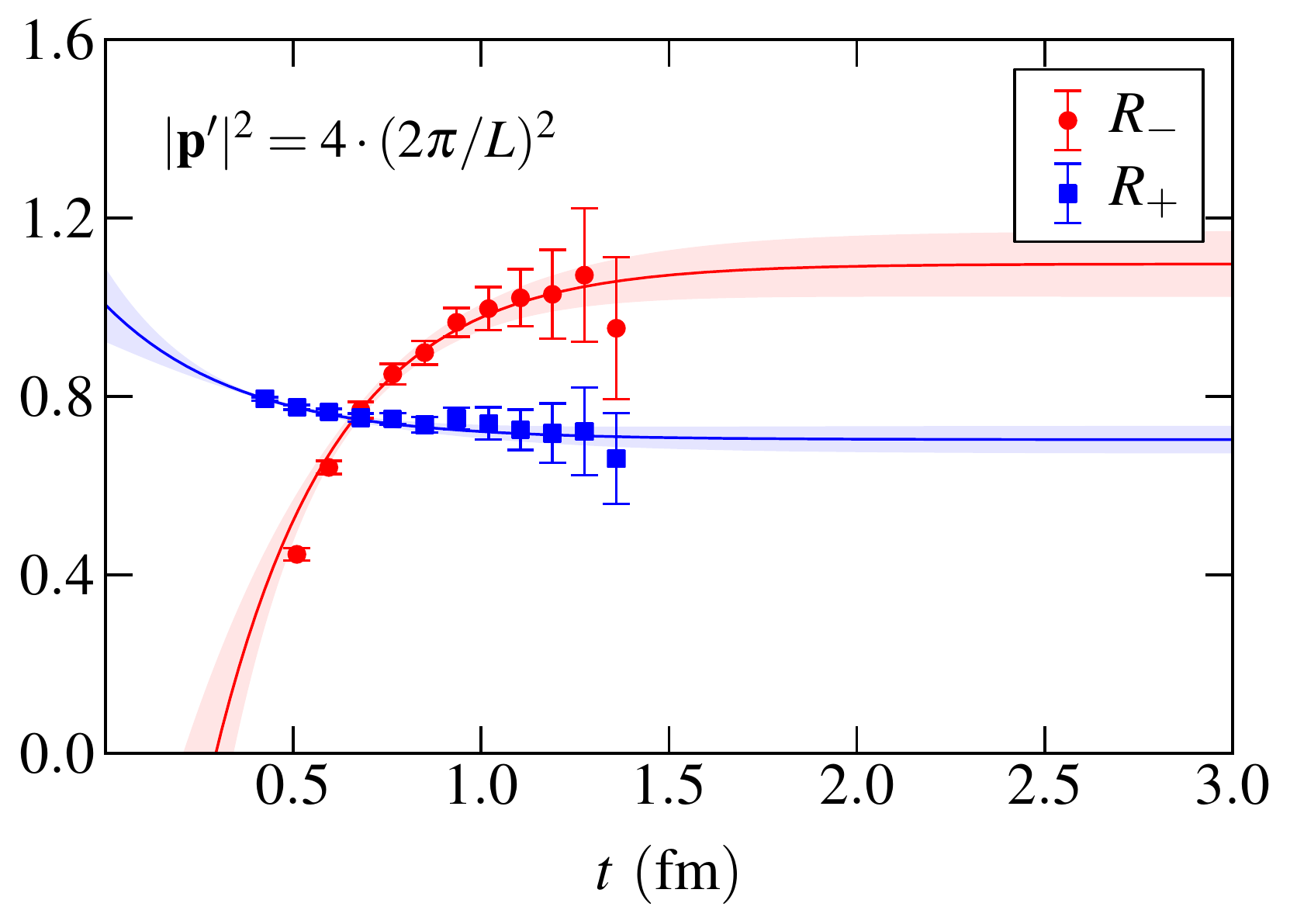}
\caption{\label{fig:tsep_dependence}Example results for $R_\pm(|\mathbf{p'}|^2, t)$,
along with fits using Eq.~(\protect\ref{eq:tdepsep}). Left panel:
$\mathtt{C54}$ data set; right panel: $\mathtt{F43}$ data set.}
\end{figure}

Our analysis uses seven different data sets $\mathtt{C14}$, ..., $\mathtt{F63}$ with parameters as shown
in Table \ref{tab:params}. We computed the ratios (\ref{eq:curlyRplus}) and (\ref{eq:curlyRminus}) for all
possible lattice momenta up to $|\mathbf{p'}|^2=9\cdot(2\pi)^2/L^2$ and for all source-sink separations in the range
from $4 \leq t/a \leq 15$ at the coarse lattice spacing and $5 \leq t/a \leq 20$ at the fine lattice spacing.
Sample results for $\mathcal{R}_\pm (|\mathbf{p'}|^2, t, t')$ are shown in Fig.~\ref{fig:ratios_L24_005_psqr4}
as a function of the current-insertion time $t'$. Note that there are plateaus in $t'$, but the non-negligible dependence
on $t$ indicates that there are still excited state-contributions. This can be seen more clearly in
Fig.~\ref{fig:tsep_dependence}, where the corresponding results for the quantities (\ref{eq:Rplus})
and (\ref{eq:Rminus}) are plotted against $t$. To isolate the ground-state contributions
(i.e., the form factors $F_\pm$), we extrapolate $R_\pm$ to $t=\infty$, allowing for excited states using the ansatz
\begin{table}
\begin{center}
\small
\begin{tabular}{cccccccccc}
\hline\hline
Set & $N_s^3\times N_t$ & $am_{s}^{(\mathrm{sea})}$  & $am_{u,d}^{(\mathrm{sea})}$
& $a$ (fm) & $am_{s}^{(\mathrm{val})}$ & $am_{u,d}^{(\mathrm{val})}$  & $m_\pi^{(\mathrm{vv})}$  & $m_{\eta_s}^{(\mathrm{vv})}$ \\
\hline
$\mathtt{C14}$ & $24^3\times64$ & $0.04$ & $0.005$ & $0.1119(17)$ & $0.04$ & $0.001$    & 245(4)   & 761(12)  \\
$\mathtt{C24}$ & $24^3\times64$ & $0.04$ & $0.005$ & $0.1119(17)$ & $0.04$ & $0.002$    & 270(4)   & 761(12)  \\
$\mathtt{C54}$ & $24^3\times64$ & $0.04$ & $0.005$ & $0.1119(17)$ & $0.04$ & $0.005$    & 336(5)   & 761(12)  \\
$\mathtt{C53}$ & $24^3\times64$ & $0.04$ & $0.005$ & $0.1119(17)$ & $0.03$ & $0.005$    & 336(5)   & 665(10)  \\
$\mathtt{F23}$ & $32^3\times64$ & $0.03$ & $0.004$ & $0.0849(12)$ & $0.03$ & $0.002$    & 227(3)   & 747(10)  \\
$\mathtt{F43}$ & $32^3\times64$ & $0.03$ & $0.004$ & $0.0849(12)$ & $0.03$ & $0.004$    & 295(4)   & 747(10)  \\
$\mathtt{F63}$ & $32^3\times64$ & $0.03$ & $0.006$ & $0.0848(17)$ & $0.03$ & $0.006$    & 352(7)   & 749(14)  \\
\hline\hline
\end{tabular}
\end{center}
\caption{\label{tab:params} Properties of the gauge-field ensembles and propagators. Here, $m_\pi^{(\mathrm{vv})}$
and $m_{\eta_s}^{(\mathrm{vv})}$ (both in units of MeV) are the masses of the pion and the pseudoscalar $s\bar{s}$
meson (without disconnected contributions), corresponding to the valence quark masses
 $am_{u,d}^{(\mathrm{val})}$ and $am_{s}^{(\mathrm{val})}$, respectively.}
\end{table}
\begin{equation}
R^{i,n}_\pm(t) = F^{i,n}_\pm + A^{i,n}_\pm \:\exp[-\delta^{i,n}_\pm\:t], \label{eq:tdepsep}
\end{equation}
where the label $i=\mathtt{C14}, ..., \mathtt{F63}$ denotes the data set, and $n$ labels the momentum squared:
$|\mathbf{p'}|^2=n\cdot(2\pi)^2/L^2$. Because the energy gap parameters $\delta^{i,n}_\pm$ in Eq.~(\ref{eq:tdepsep})
are positive by definition, we rewrite them as $\delta^{i,n}_\pm = \exp(l^{i,n}_\pm) \cdot (1\: {\rm GeV})$,
and use $l^{i,n}_\pm$ along with $F^{i,n}_\pm$ and $A^{i,n}_\pm$ as the fit parameters. Where necessary,
we exclude a few points at the smallest $t$ from the fits to get good $\chi^2/{\rm d.o.f}$.
At fixed momentum $n$, we perform the fits simultaneously for the different data sets $i$.
Since $2\pi/L$ (in GeV) is equal within uncertainties for the coarse and fine lattice spacings,
we know (from prior studies of the hadron spectrum on the lattice) that the physical energy gaps must be
of similar size for the different data sets $i$. With this knowledge, we use Bayesian constraints
that limit differences $|l^{i,n}_\pm-l^{j,n}_\pm|$ to reasonable values
(details will be given in Ref.~\cite{paper}), which improves the stability of the fits.

\begin{figure}
\includegraphics[width=0.48\linewidth]{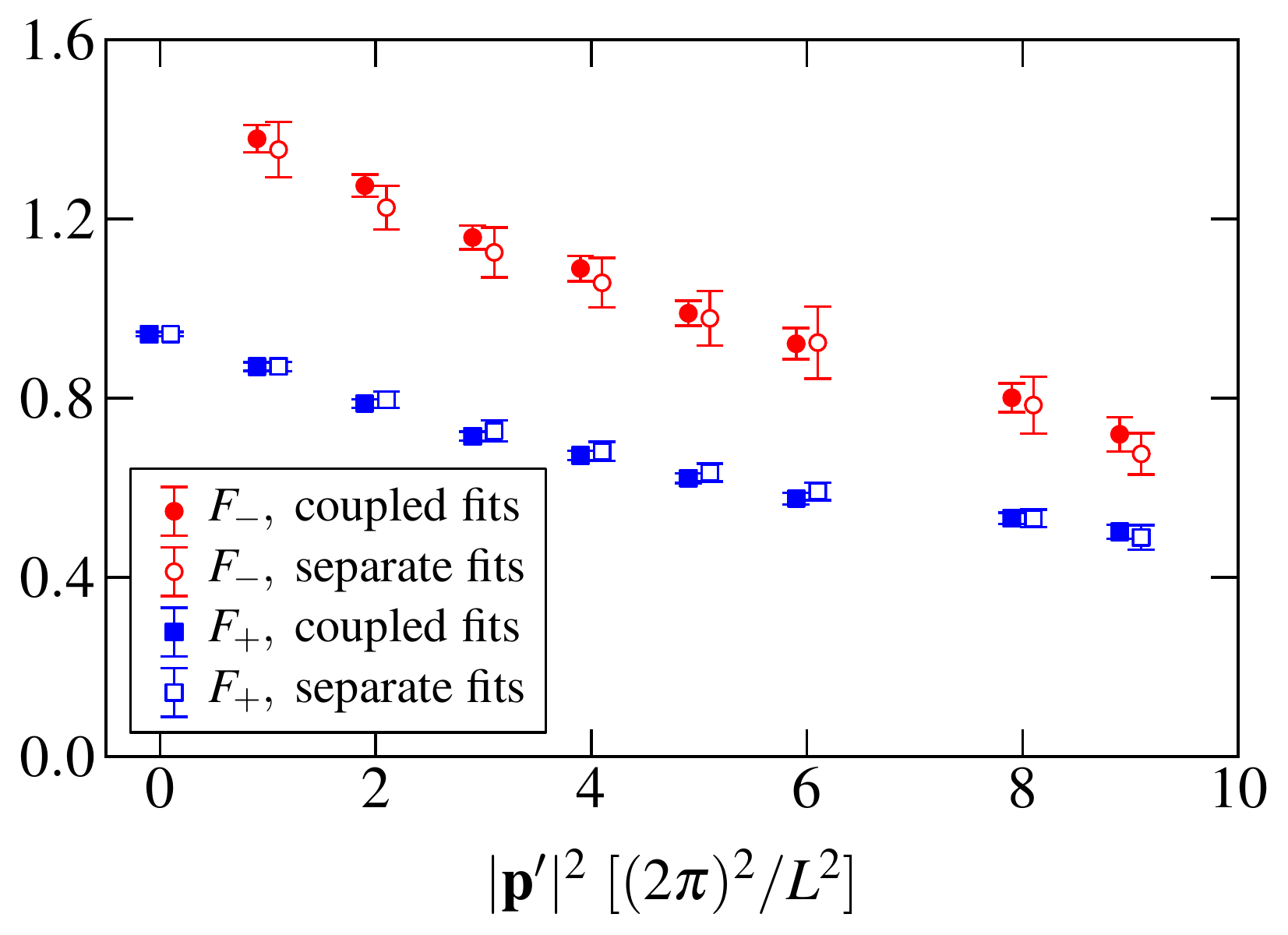}
\hfill\includegraphics[width=0.48\linewidth]{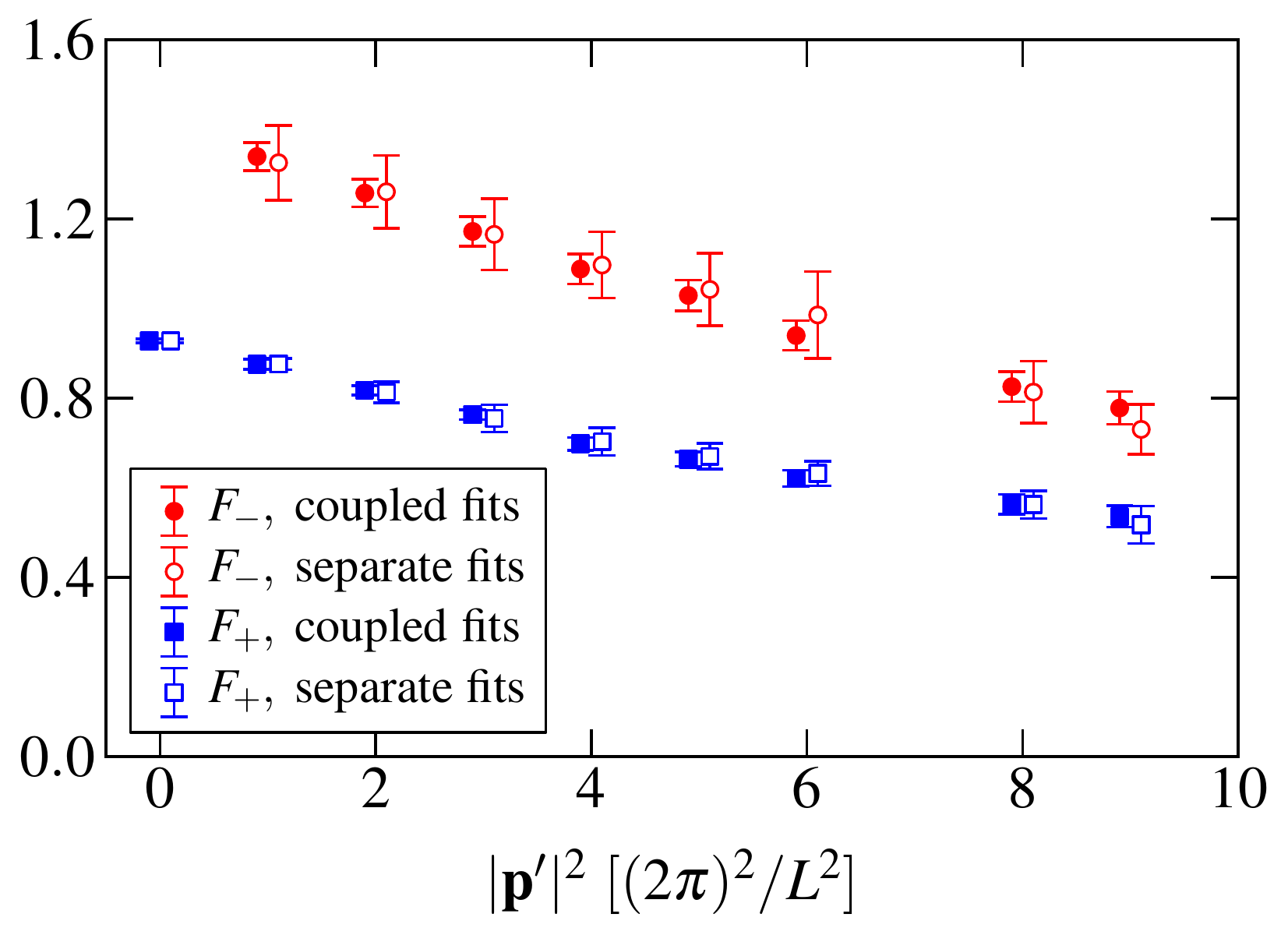}
\caption{\label{fig:resultsinfinitesepcoupledvsseparate}Fit results for $F^{i,n}_\pm$
from the data sets $i=\mathtt{C54}$ (left) and $i=\mathtt{F43}$ (right). Open symbols show
results obtained using Eq.~(\protect\ref{eq:tdepsep}) with separate energy gap parameters
$l^{i,n}_+$ and $l^{i,n}_-$. Filled symbols show the results from coupled fits with
$l^{i,n}_+=l^{i,n}_-=l^{i,n}$. The points are offset horizontally for clarity.}
\end{figure}

Having performed these fits, we noted that at each momentum $n$ and data set $i$, the energy gap parameters
$l^{i,n}_+$ and $l^{i,n}_-$ returned from the fit were equal within uncertainties. This is expected as
long as the relevant excited states have non-zero matrix elements in both $R_+$ and $R_-$. We therefore performed
new fits with shared energy gap parameters $l^{i,n}_+=l^{i,n}_-=l^{i,n}$. These new, coupled fits had values
of $\chi^2/{\rm d.o.f}$ as good as or better than the separate fits, and the extracted form factors were consistent
with those from separate fits (see Fig.~\ref{fig:resultsinfinitesepcoupledvsseparate}), so we use $F^{i,n}_\pm$
from the coupled fits in the further analysis. To estimate the systematic uncertainty resulting from the choice
of the $t_{\rm min}$'s, we compute the shifts in $F^{i,n}_\pm$ when increasing all $t_{\rm min}$'s by one unit,
and we add these shifts in quadrature to the statistical uncertainties.

\begin{figure}
\includegraphics[width=0.48\linewidth]{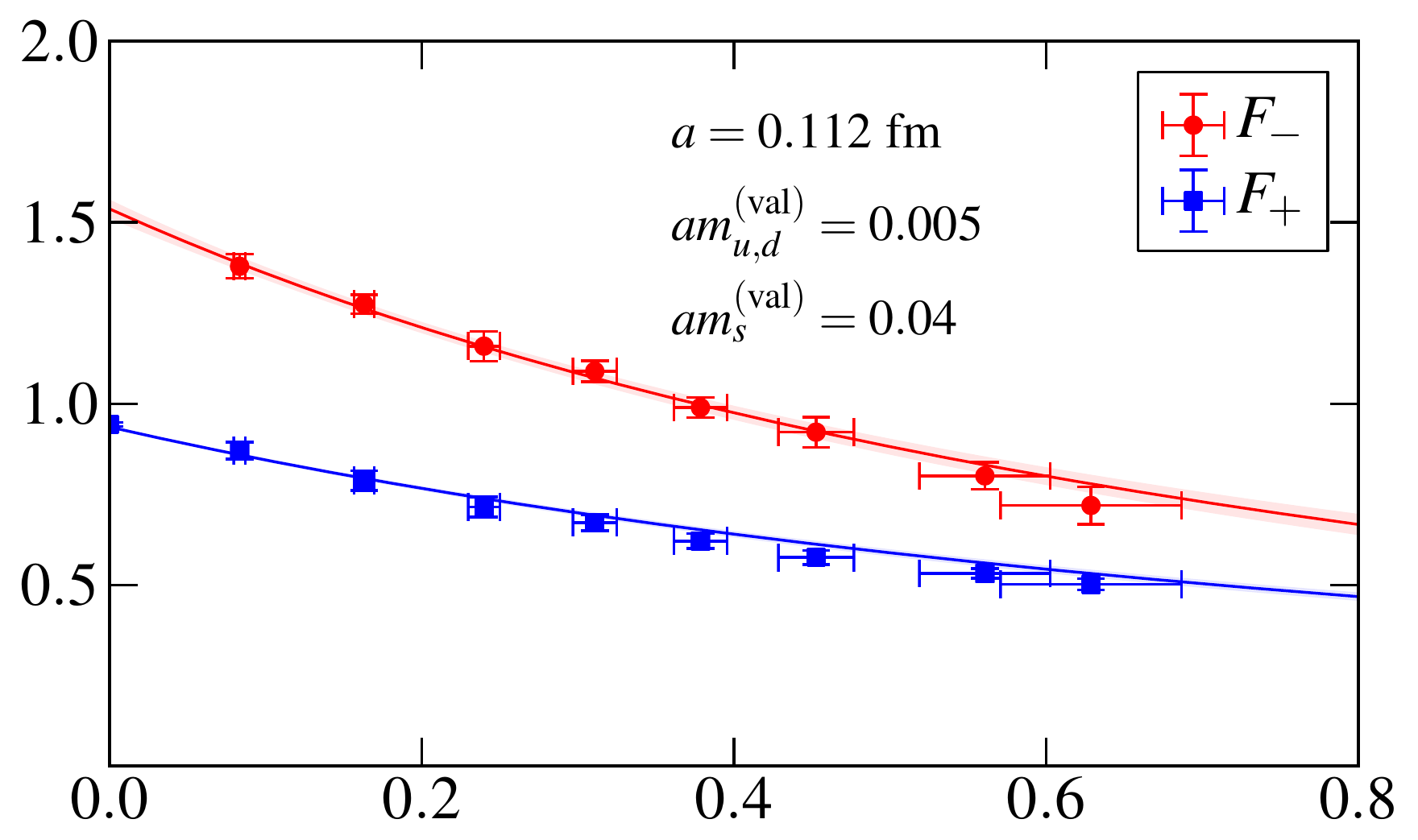}
\hfill\includegraphics[width=0.48\linewidth]{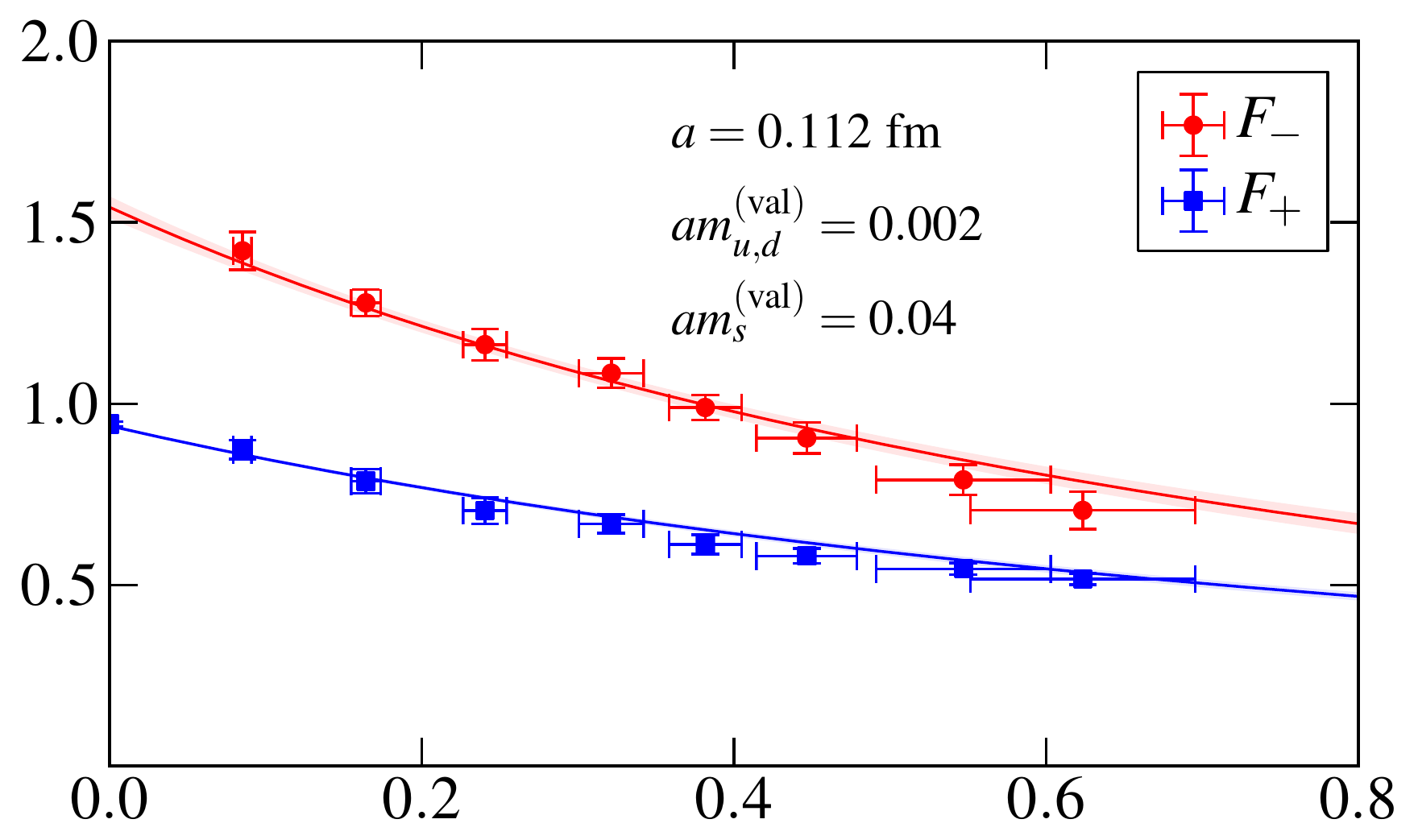} \\
\includegraphics[width=0.48\linewidth]{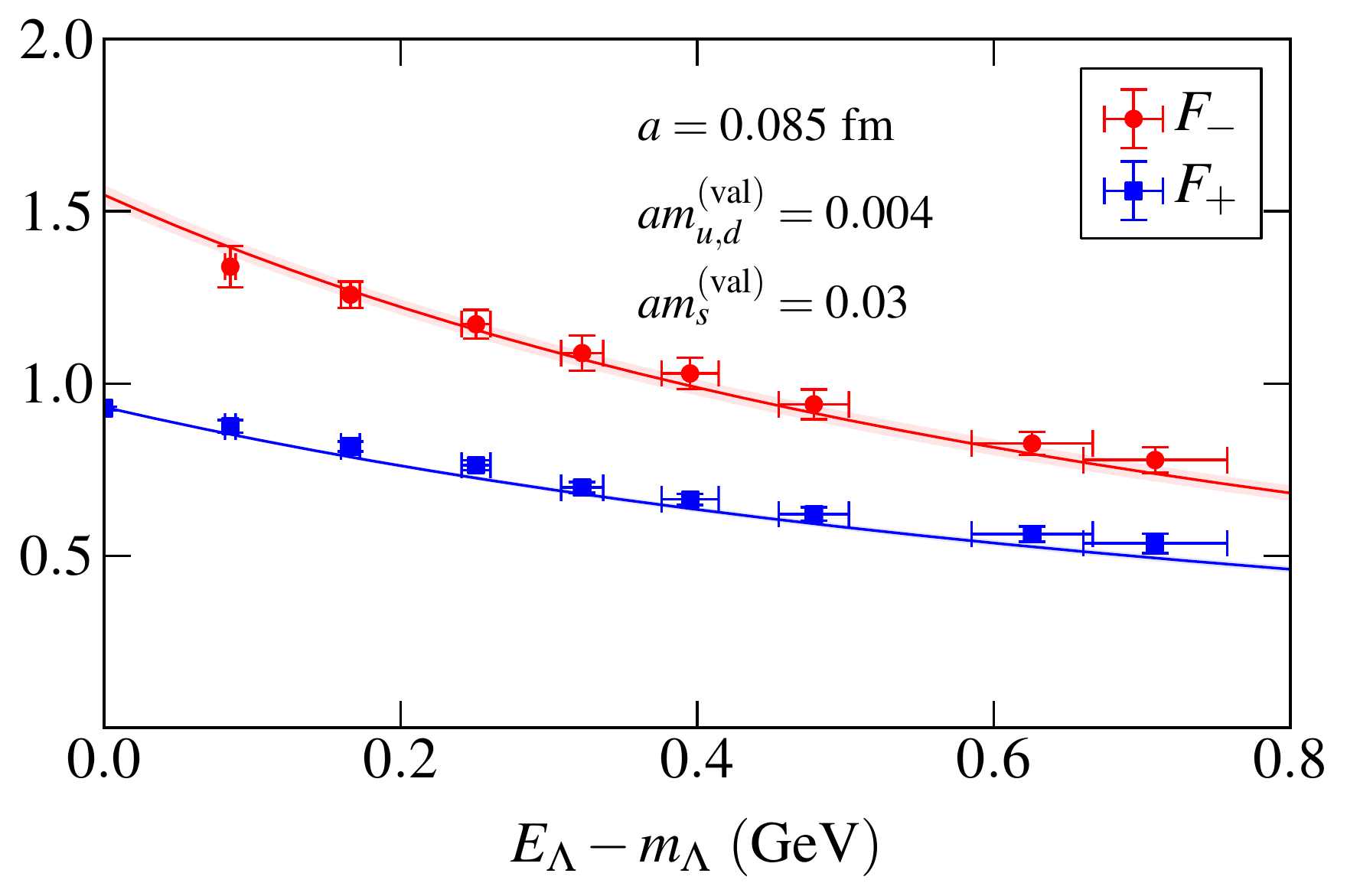}
\hfill\includegraphics[width=0.48\linewidth]{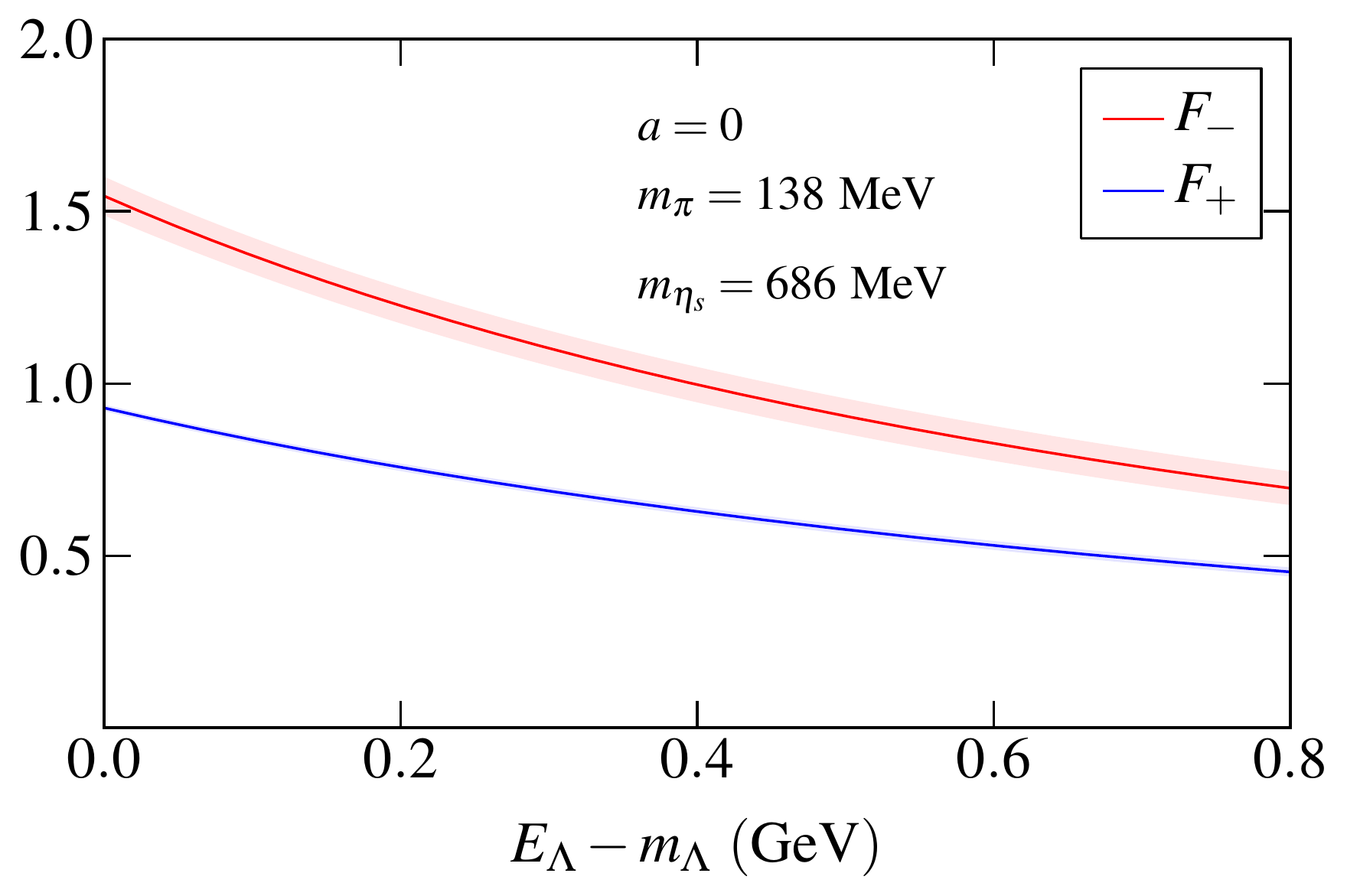}
\caption{\label{fig:qsqrasqrextrapall}Preliminary fits of the form factor data for $F_+$ and $F_-$
using Eq.~(\protect\ref{eq:dipole}). The fits includes all seven data sets
(see Table \protect\ref{tab:params}), but only three data sets are shown to save space.
The bottom-right plot shows the fitted functions evaluated in the continuum limit and at the
physical values of the light and strange-quark masses.}
\end{figure}

The last step of the analysis is to interpolate the dependence of the form factors on $p'\cdot v=E_\Lambda$ using
a suitable smooth function, and extrapolate to the continuum limit and the physical $u$, $d$, and $s$-quark masses.
Since $E_\Lambda$ depends strongly on the quark masses, it is better to consider the form factors on the lattice as
functions of $E_\Lambda-M_\Lambda$. At the present level of statistical precision, and for the energy range
considered here, we find that generalized dipole fits of the form
\begin{eqnarray}
 F_\pm^{i,n} &=& \frac{N_\pm}{(X_\pm^i+E_\Lambda^{i,n}-m_\Lambda^i)^2}\cdot [1 + d_\pm (a^i E_\Lambda^{i,n})^2], \label{eq:dipole} \\
X_\pm^i &=& X_\pm + c_{l,\pm}\cdot \left[ (m_\pi^i)^2-(m_\pi^{{\rm phys}})^2\right]
+ c_{s,\pm} \cdot \left[ (m_{\eta_s}^i)^2-(m_{\eta_s}^{{\rm phys}})^2 \right], \label{eq:polemqdep}
\end{eqnarray}
with parameters $N_\pm$, $X_\pm$, $d_\pm$, $c_{l,\pm}$, and $c_{s,\pm}$ describe our data well
($\chi^2/{\rm dof}=0.59$ for $F_+$, and $\chi^2/{\rm dof}=0.43$ for $F_-$). In the continuum limit and for the physical
light and strange-quark masses (we use $m_{\eta_s}^{{\rm phys}}=686$ MeV \cite{Davies:2009tsa}), these functions reduce
to $F_\pm = N_\pm/(X_\pm+E_\Lambda-m_\Lambda)^2$, which contain only the fit parameters $N_\pm$ and $X_\pm$.
Our preliminary results for these parameters are given in Table \ref{tab:dipolefitresults}. Plots of the fitted
functions are shown in Fig.~\ref{fig:qsqrasqrextrapall}.

\begin{table}
\begin{center}
\small
\begin{tabular}{ccc}
\hline\hline
Parameter & \hspace{1ex} & Result \\
\hline
$N_+$  && $3.188 \pm 0.268$ ${\rm GeV}^2$  \\
$X_+$  && $1.852 \pm 0.074$ ${\rm GeV}^{\phantom{2}}$  \\
$N_-$  && $4.124 \pm 0.750$ ${\rm GeV}^2$ \\
$X_-$  && $1.634 \pm 0.144$ ${\rm GeV}^{\phantom{2}}$  \\
\hline\hline
\end{tabular}
\end{center}
\caption{\label{tab:dipolefitresults} Preliminary results for $N_\pm$ and $X_\pm$ from fits using Eq.~(\protect\ref{eq:dipole}).}
\end{table}

\section{Outlook}

We have performed the first lattice calculation of the $\Lambda_Q \to \Lambda$ form factors
$F_+=F_1+F_2$ and $F_-=F_1-F_2$. Using a ratio technique with a wide range of source-sink separations,
we have achieved a high level of statistical precision. The dominant systematic uncertainties in our results
are associated with the use of one-loop perturbative current matching, finite-volume effects, the naive linear
extrapolations in the light-quark masses, and the continuum extrapolations. We estimate that the total
systematic uncertainty is below 10\%; more details will be given in Ref.~\cite{paper}. There, we will also
present results for the differential branching fraction of $\Lambda_b \to \Lambda\mu^+\mu^-$.

\noindent \textbf{Acknowledgments}: This work is supported by the U.S.~Department of Energy under cooperative
research agreement Contract Number DE-FG02-94ER40818. Numerical computations were performed using resources at
NERSC (funded by DOE grant number DE-AC02-05CH11231) and XSEDE resources at NICS (funded by NSF grant number OCI-1053575).

\linespread{1.05}

\end{document}